\documentclass[journal,transmag]{IEEEtran}
 \pdfoutput=1
 \usepackage{dcolumn}
 \usepackage{graphicx,color}
 \usepackage[cmex10]{amsmath} 
 \usepackage{amssymb} 
 \usepackage{array}
 \usepackage{cite}\usepackage[utf8]{inputenc}
 \usepackage{comment}

\title{Architectural considerations in the design of a third-generation superconducting quantum annealing processor}
\author{\parbox{\linewidth}{\centering Kelly~Boothby, Colin~Enderud, Trevor~Lanting, Reza~Molavi, Nicholas~Tsai, Mark~H.~Volkmann, Fabio~Altomare, Mohammad~H.~Amin, Michael~Babcock, Andrew~J.~Berkley, Catia~Baron~Aznar, Martin~Boschnak, Holly~Christiani, Sara~Ejtemaee, Bram~Evert, Matthew~Gullen, Markus~Hager, Richard~Harris, Emile~Hoskinson, Jeremy~P.~Hilton, Kais~Jooya, Ann~Huang, Mark~W.~Johnson, Andrew~D.~King, Eric~Ladizinsky, Ryan~Li, Allison~MacDonald, Teresa~Medina~Fernandez, Richard~Neufeld, Mana~Norouzpour, Travis~Oh, Isil~Ozfidan, Paul~Paddon, Ilya~Perminov, Gabriel~Poulin-Lamarre, Thomas~Prescott, Jack~Raymond, Mauricio~Reis, Chris~Rich, Aidan~Roy, Hossein~Sadeghi~Esfahani, Yuki~Sato, Ben~Sheldan, Anatoly~Smirnov, Loren~J.~Swenson, Jed~Whittaker, Jason~Yao, Alexander~Yarovoy, and~Paul~I.~Bunyk\\
\vspace{1em}
\emph{in memory of Paul Bunyk, our dear friend}
}%
\thanks{K.~Boothby, C.~Enderud, T.~Lanting, R.~Molavi, N.~Tsai, M.~H.~Volkmann, F.~Altomare, M.~H.~Amin, M.~Babcock, A.~J.~Berkley, C.~Baron~Aznar, M.~Boschnak, H.~Christiani, S.~Ejtemaee, M.~Gullen, M.~Hager, R.~Harris, E.~Hoskinson, A.~Huang, M.~W.~Johnson, A.~King, E.~Ladizinsky, R.~Li, A.~MacDonald, T.~Medina~Fernandez, R.~Neufeld, M.~Norouzpour, T.~Oh, P.~Paddon, G.~Poulin-Lamarre, T.~Prescott, J.~Raymond, M.~Reis, C.~Rich, H.~Sadeghi~Esfahani, Y.~Sato, B.~Sheldan, A.~Smirnov, L.~Swenson, J.~Whittaker, J.~Yao, and A.~Yarovoy are with D-Wave Systems, Inc., Burnaby, BC, V5G 4M9, Canada (e-mail: boothby@dwavesys.com)}%
\thanks{J. P. Hilton is with Google AI Quantum, Goleta, CA, 93111, USA}%
\thanks{B. Evert is with Rigetti Computing, Berkeley, CA, 94710, USA}%
\thanks{L. Swenson is with AWS  Center  for  Quantum  Computing,  Pasadena,  CA  91125,  USA}%
}
\def\fdac{$\Phi$-DAC}

\begin{document}

\maketitle
\begin{abstract}
Early generations of superconducting quantum annealing processors have provided a valuable platform for studying the performance of a scalable quantum computing technology. These studies have directly informed our approach to the design of the next-generation processor. Our design priorities for this generation include an increase in per-qubit connectivity, a problem Hamiltonian energy scale similar to previous generations, reduced Hamiltonian specification errors, and an increase in the processor scale that also leaves programming and readout times fixed or reduced. Here we discuss the specific innovations that resulted in a processor architecture that satisfies these design priorities.
\end{abstract}

\section{Introduction}

\begin{figure}[b]
 \begin{center}
  \includegraphics[width=.45\textwidth]{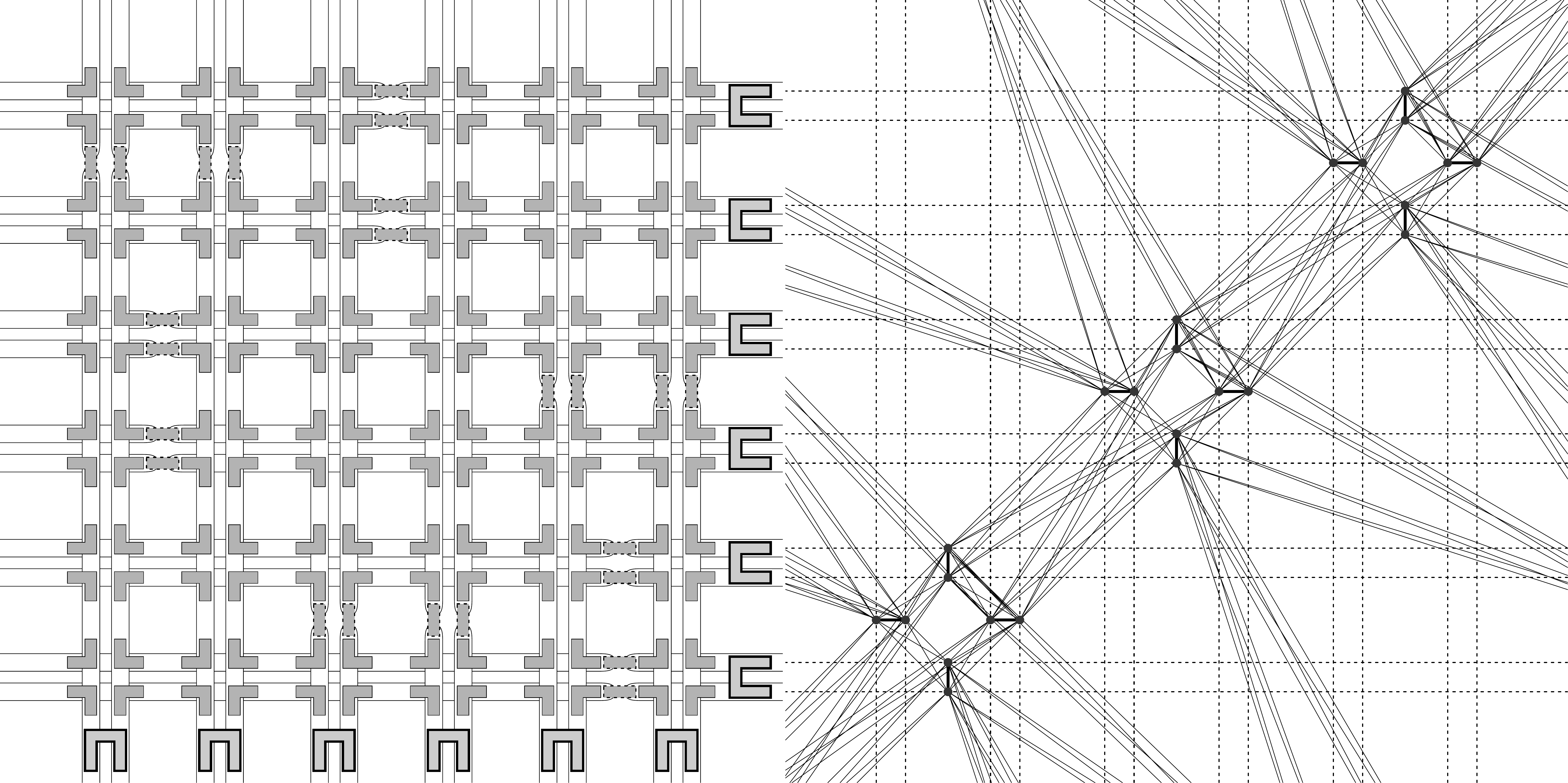}
 \end{center}
 \caption{Tile architecture.  \textit{left}   Layout sketch of qubits and couplers.  Qubit bodies are represented by elongated loops, not fully contained within the tile.  The qubits are arranged in pairs, with odd couplers (thick outline, U-shaped objects) between the members of a pair.  External couplers (dashed outline rectangles) are distributed throughout the tile to connect rows and columns of parallel qubits.  Internal couplers (L-shaped objects) connect perpendicular qubits. \textit{right} Graph representation.  Each unit tile contains 24 nodes (circles) which are connected in odd pairs (heavy lines),  have external couplers to neighboring tiles (dashed lines), and internal couplers within the tile (fine lines).
 }\label{fig:unittile}
\end{figure}

Concurrently with customer and internal evaluation of our previous-generation {\sc D-Wave Two\textsuperscript{TM} }\cite{Bunyk2014} processors, we launched an effort to further improve upon the state of the art in the design of quantum processing units (QPU), the superconducting integrated circuit at the heart of our quantum computers. This effort culminated in our current {\sc Advantage\textsuperscript{TM} } quantum computer\cite{Boothby2020}. 

It is clear that limited connectivity in our previous processors, while allowing impressive proof-of-concept demonstrations\cite{Harris2018, King2018}, requires long-\emph{chain} \emph{embeddings}\cite{Choi2008, Choi2010}, for many commercially relevant problems.  An embedding is a representation of a desired Hamiltonian $H$ which uses strong antiferromagnetic couplings to produce chains of qubits which represent the variables of $H$.  Where long-chain embeddings can be fit into the processor qubit count, the corresponding decrease in chain (logical qubit) energy scales commonly results in a sharp decline of solution success probabilities.

Logical qubit performance is informed by the integer length of the physical qubit chain underlying them, as well as the energy scales of the chained individual qubits and couplings. Embedded qubit chain lengths follow from the QPU graph topology, crudely characterizable by its \emph{connectivity}, a term we use to describe the density and variety of edges in the graph. Graphs with better connectivity lead to shorter chain lengths for relevant problems. For the purposes of the following discussion, connectivity may be read as a synonym for vertex (qubit) degree.

Increased qubit energy scales can be achieved by decreasing the qubit's physical size. In our QPUs, the physical qubits and couplers are laid out in a multilayer fabric, interleaved with their control circuitry. Physical qubit size is then set by the size of the control circuitry required to operate the qubit and attached couplers.

Increased connectivity can be achieved, in turn, by increasing the density of devices. If we fix qubit size (perhaps to fix energy scales), the required areal device density increases with connectivity. Conversely, if we fix the areal device density, qubit size grows as connectivity increases.

Better qubit technology was needed, with the specific goals of increasing connectivity, qubit energy scales, or some combination of both. A more favorable design space is accessible by decreasing the size of our control circuitry, which translates to reducing the footprint of our digital-to-analog persistent current control devices (\fdac) \cite{Johnson2010, Bunyk2014}.

To this end, we made significant changes to the fabrication technology.  Along the lines suggested in Section III.D of our previous QPU architecture report \cite{Bunyk2014}, these features enabled us to shrink the single \fdac\ areal footprint by a factor of 3. Moreover, within the new footprint we managed to fit not two, but four \fdac\ stages, substantially increasing programming precision while maintaining comparable \fdac\ programming time.

For our current {\sc Advantage} offering we opted to use the available gains for maximizing connectivity, while maintaining energy scales similar to our previous generation.  Detailed experimental studies of the trade-off between connectivity and energy scale using problem performance are under way and we plan to discuss this in a future publication.

In addition, we decreased control errors using better precision on \fdac s, reduced parasitic cross-talks by taking advantage of more favorable design rules, and made calibration improvements by means of concurrency, partially enabled by a new readout system. While each qubit looks conceptually and parametrically similar to those composing the {\sc D-Wave Two} {\em Chimera} topology, it is connected not to 6, but to 15 others (away from the edges of the graph).

Section~\ref{sec:topology} introduces our new topology, enabled by much greater flexibility in arranging highly connected qubits. In Section~\ref{sec:energy}, we discuss the energy scale metric we used to optimize qubit design. Section \ref{sec:control} discusses the on-chip control circuitry, including a new \fdac\ addressing scheme. Section~\ref{sec:readout} introduces a new readout scheme developed for {\sc Advantage}. We discuss the advantages of design modularity in Section~\ref{sec:floorplan} before presenting our concluding remarks in Section~\ref{sec:conclusion}. 

\section{Topology}
\label{sec:topology}
The previous-generation architecture was comprised of qubits in a regular array of square tiles, each containing 8 qubits, and each qubit connected to (up to) 6 couplers.  The current generation is a significant departure from the previous design, where the tiles contain a grid of 144 \emph{internal} couplers (those between perpendicular qubit pairs), 24 \emph{external} couplers (those between colinear qubits) and 12 \emph{odd} couplers (discussed below).  The qubits in this arrangement are not constrained to a single tile, rather, each qubit straddles the boundary between two adjacent tiles; and each qubit is connected to up to 15 couplers. 

\subsection{Odd Coupler}

The current generation features a novel coupler arrangement; namely the \emph{odd} coupler, as described in \cite{Boothby2020}.  This type of coupler is enabled by a new plaquette design --- by arranging qubits in pairs,  we find efficiency both in the control plaquettes and in the opportunity to add a coupler between the paired qubits.  See Figure~\ref{fig:unittile} for a bird's-eye view of the tile layout, showing the placement of odd couplers.  These couplers occupy a portion of the qubit that was otherwise uncoupled, so they incur negligible cost in terms of real estate and energy scale.  Moreover, these couplers provide a mechanism for solving problems with length-3 cycles without the need for chains, which was not possible in previous topologies.

\section{Energy Scale}
\label{sec:energy}

The energy Hamiltonian in a quantum annealer can be expressed as:

$H = A(s)\big(-\sum_i \sigma_x^i \big) + B(s)\big( \sum_i h_i \sigma_z^i +\sum_{i,j} J_{ij} \sigma_z^i \sigma_z^j \big)$
where the \emph{annealing schedule}, $s \in [0, 1]$, is a function of time $s(t)$ and parameters $A$ and $B$ are dependent on $s$. During an annealing process where $s$ increases from zero to one, the tunneling energy of single qubit, represented by $A$, monotonically decreases while the coupling energy between qubits, represented by $B$, increases, forming the problem $H$ energy landscape. A useful energy scale when evaluating potential architectures is at the one-dimensional quantum critical point (QCP), where $E_{QCP} := A(s) = B(s)$ (this is the point at which an infinitely long one-dimension chain of qubits with coupling $|J_{i,i+1}|=1$ makes a transition from a paramagnetic to an ordered state). A key design goal is to maximize $E_{QCP}$ and in particular, keep it well above the temperature of the bath to which the qubits are coupled, $ E_{QCP} = A_{QCP} \gg k_{b}T$, to avoid single qubit thermal excitations.

Both $A(s)$ and $B(s)$ are functions of the inductance, capacitance, and critical current of the physical flux qubits.  $B(s) = M_{AFM}|I^p_q|^2$ depends on the mutual inductance between pairs of flux qubits $M_{AFM}$ and the persistent current $I^p_q(s)$ that can flow in the qubit.  While the flux qubit is best represented by a distributed circuit model with several devices attached, one can usually reduce it to an effective model with lumped element inductors, capacitors, and Josephson junctions.  To ensure quantum mechanical effects are dominant throughout the anneal, it is crucial to carefully design and engineer the qubit physical parameters.

While closed-form analytical expressions allow a qualitative study of how different devices impact qubit impedance, it becomes prohibitively less efficient for larger arrays of qubits with coupling to many neighboring qubits. The impact of magnetic shielding surrounding qubits and capacitive coupling into neighboring structures are harder to assess using compact models. To estimate energy scale and QCP for qubits with a connectivity degree 15, a combination of 3-D electromagnetic (EM) simulators and circuit simulator (SPICE) are used to extract qubit physical parameters. A two-fluid model in the EM solver captures kinetic inductance of superconducting qubit leads. Using a similar method, $M_{AFM}$ between pair of qubits is extracted for a pair of qubits. Using the extracted inductance and capacitance and employing a quantum mechanical model for radio freqency superconducting quantum interference devices (RF-SQUIDs) we were able to estimate the tunneling energy $A(s)$, persistent current, $|I^p_q|$, and eventually $E_{QCP}$ of the qubit. The above mentioned procedure is iterated for a variety of qubit geometries and device arrangements to arrive at optimal configuration for maximum QCP. Qubit-coupler assembly, in particular, is optimized to provide maximum magnetic coupling efficiency with minimal parasitic capacitance overhead.

\section{On-Chip Control}
\label{sec:control}

\subsection{DAC Addressing}

\begin{figure}
 \begin{center}
 \begin{tabular}{c}
  \includegraphics[scale=1]{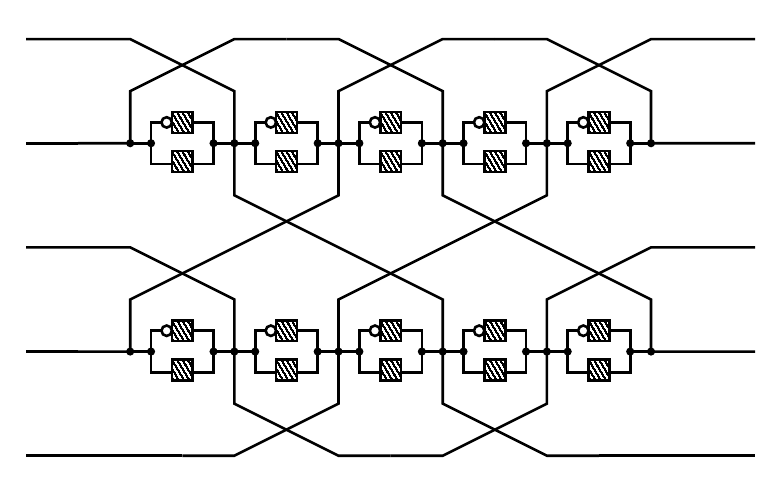}
 \end{tabular}
 \end{center}
 \caption{Demonstration circuit for a single \emph{power domain} of five address lines used to control 20 DAC stages.  Shaded boxes represent individual DAC stages; power line not shown.  Multiple power domains may be implemented by either repetition of this pattern, or by an interleaving pattern.  Negation symbol (small white circle) indicates a reversal of polarity. }\label{fig:braidunit}
\end{figure}

In the previous generation, individual DAC stages were similarly controlled by three lines known as \emph{address}, \emph{trigger}, and \emph{power}.  The architecture for these stages was a so-called XYZ addressing scheme --- the address and trigger lines are used analogously to a traditional matrix multiplexing scheme used in semiconductor electronics; an $x \times y$ grid of address and trigger lines is repeated in a supergrid with $z$ cells, each associated with a single power line.  With a single line of each type, we are able to control two DAC stages --- one where the address/trigger lines have the same polarity, the other where the polarity of one is reversed.  Thus, with $x$ address lines, $y$ trigger lines, and $z$ power lines, we could control $2xyz$ DAC stages.

It is observed that the waveforms for programming the address and trigger lines are identical, and there is no a-priori reason to keep these wires as separate groups.  The current generation uses a multiplexing scheme similar to \emph{Charlieplexing}\cite{charlieplexing}, which allows all pairs of address lines to be used for DAC addressing.  For a fixed number of power lines (we set $z=1$ for simplicity), the number of controlled DACs is maximized when $x=y$, where we can control $2x^2$ DACs at the cost of $x+y=2x$ control lines.  If we were to mix the address and trigger roles, the same $2x$ lines can control $4x^2-2x$ DACs --- nearly twice as many (in practice, an odd number of lines is preferred).  We achieve this through a braided wiring pattern.  See Figure~\ref{fig:braidunit} for a simple demonstration of the braiding pattern.  The $P_{16}$ contains 401,408 DAC stages, controlled by 57 address lines and 128 power lines.

\subsection{DAC Dynamic Range}

\begin{figure}
 \begin{center}
 \begin{tabular}{c}
   \includegraphics[width=0.25\textwidth]{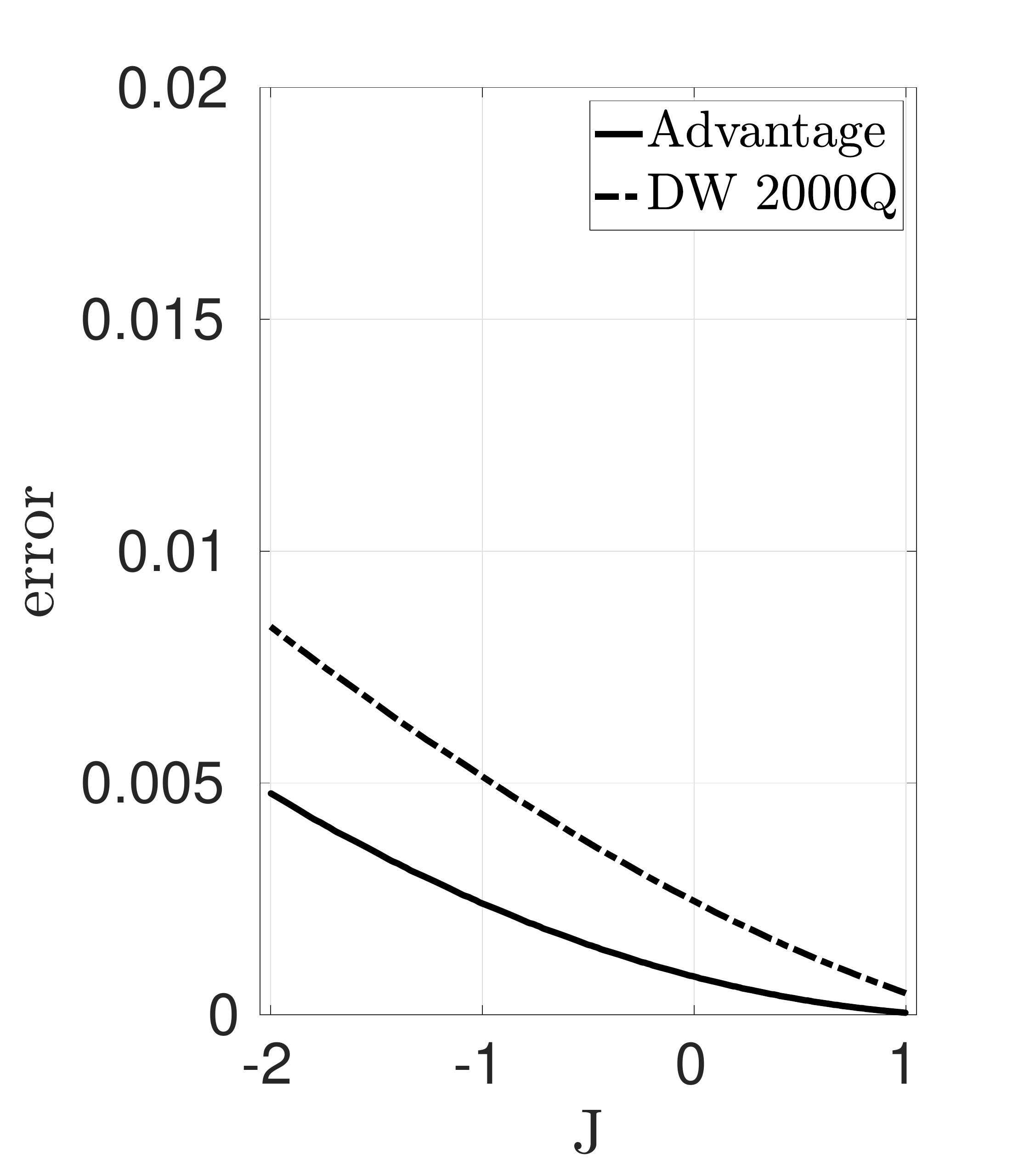}
   \includegraphics[width=0.25\textwidth]{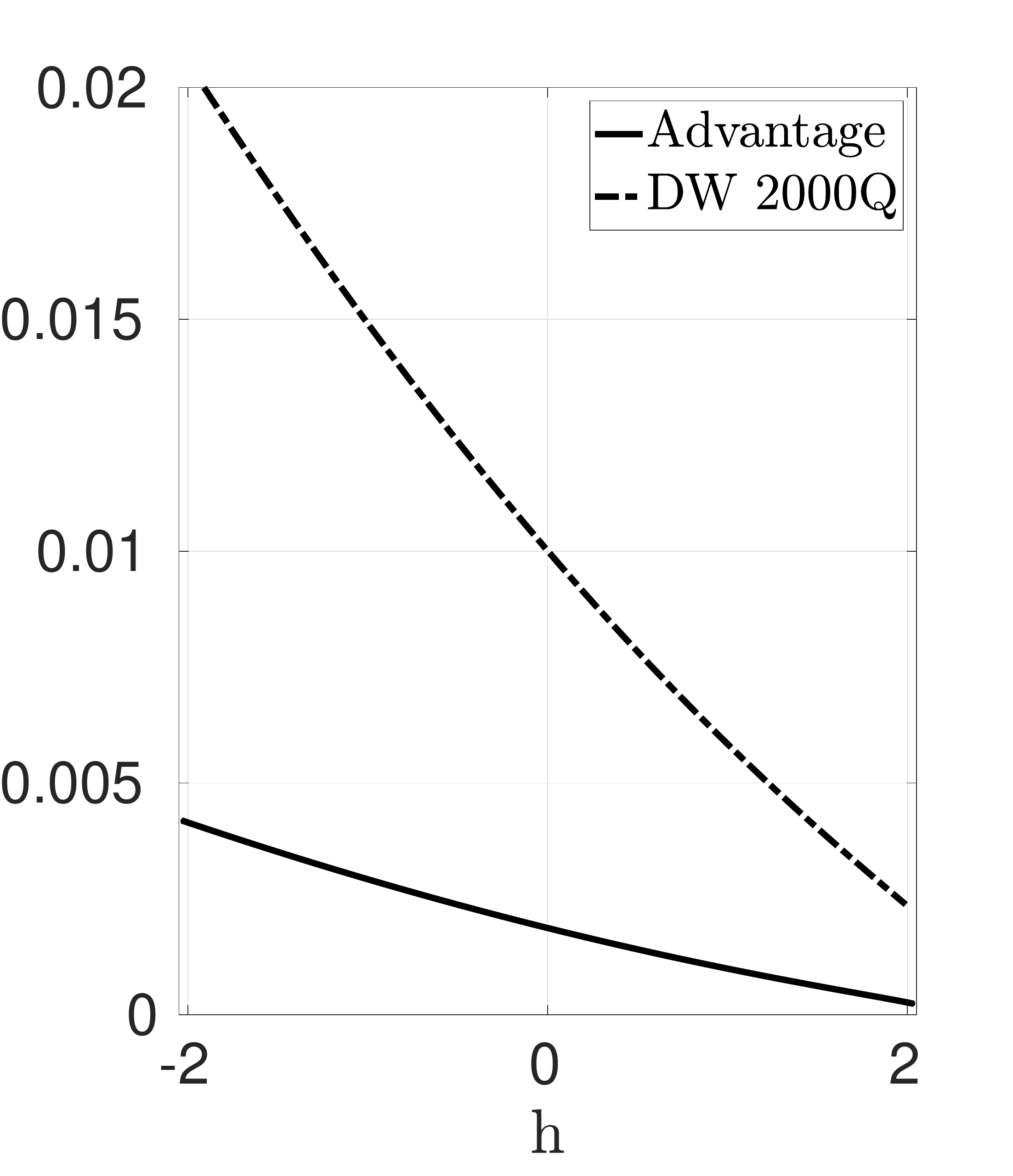}
 \end{tabular}
 \end{center}
 \caption{We increased the dynamic range of the control DACs to reduce the Hamiltonian specification error associated with DAC quantization. These plots compare this quantization error in the {\sc Advantage} architecture to that in the previous generation DW2000Q architecture.}\label{fig:ICE}
\end{figure}

A key goal for the {\sc Advantage} architecture is reducing Hamiltonian specification errors $\delta h, \delta J$ from previous generations. The DAC dynamic range was increased for several DAC types to reduce the errors associated with DAC quantization. Figure~\ref{fig:ICE} shows a comparison of this quantization error for the {\sc Advantage} architecture and the previous generation. Quantization errors were reduced by over $2$ times the previous generation and we achieved a contribution from DAC quantization of $\delta h, \delta J < 0.01$.

\section{Readout}
\label{sec:readout}

At the end of the annealing schedule the tunneling energy $A(s)$ is negligible and the qubits have each localized into one of the two flux eigenstates with macroscopic persistent current circulating in one of two directions, corresponding to these two eigenstates. To read out the processor we do a high fidelity measurement of the state of every qubit. There are two components to the readout circuitry. The first is a flux sensitive shift register that moves data from the interior of the processor to the perimeter. Around the perimeter we place an array of superconducting microresonators to read out the data in the shift register~\cite{Whittaker2016}.

\begin{figure}
 \begin{center}
 \begin{tabular}{c}
  \includegraphics[scale=.6]{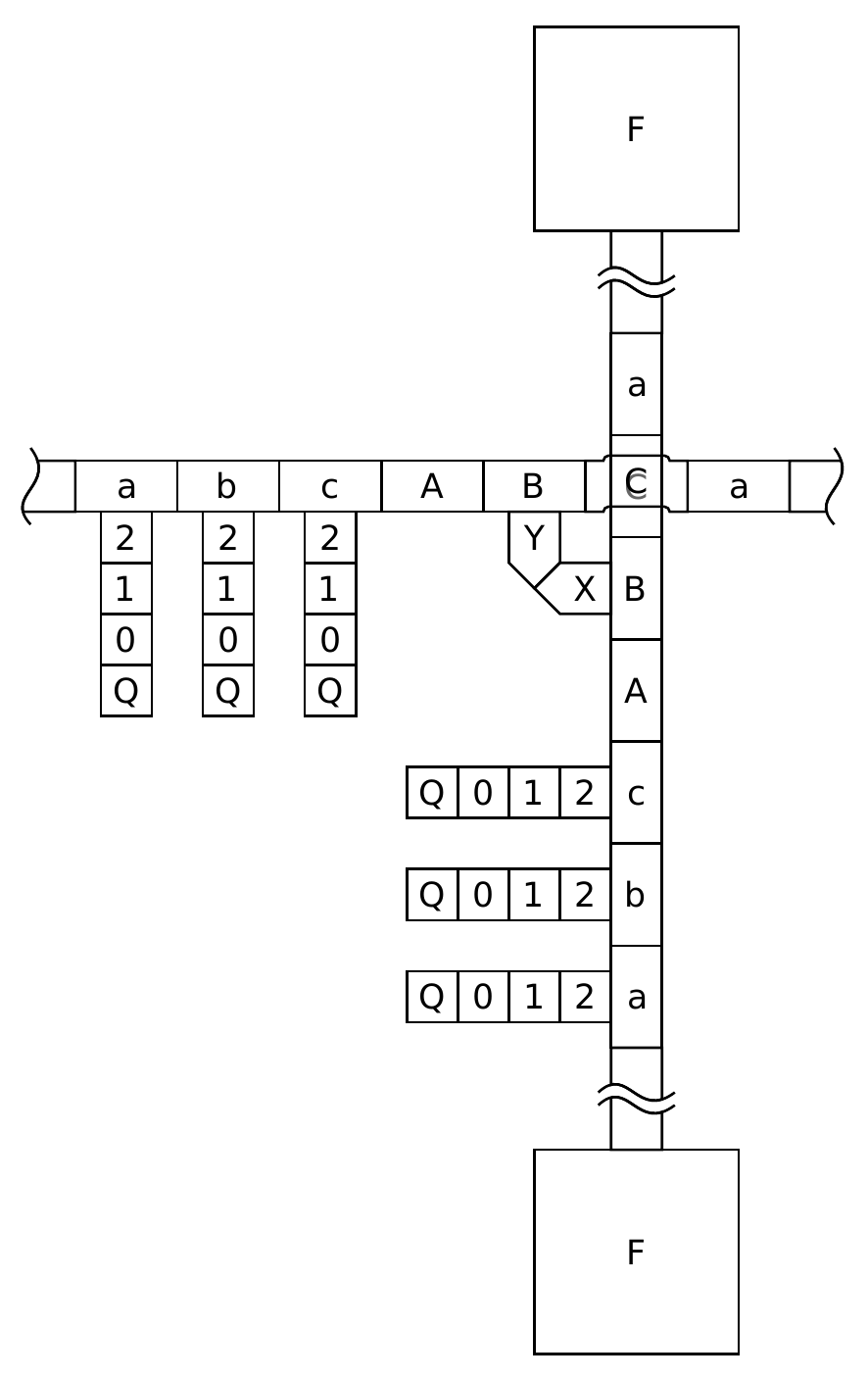}
 \end{tabular}
 \end{center}
 \caption{Shift register fabric as connected to FASTRs (labeled F) and qubits (labeled Q).  QFP stages are shown with associated clock line labels ($0, 1, 2, a, b, c, A, B, C, X, Y$).  Drop-shadow effect on $C$ signifies two crossing stages with the same clock line.  Actual clock line labeling may differ between versions.}\label{fig:srs}
\end{figure}
\subsection{Shift Register Fabric}

We use a \emph{quantum flux parametron} (QFP)-based shift register to move data from the qubits along linear horizontal and vertical tracks to the perimeter of the processor (see Figure~\ref{fig:srs}). In contrast to the previous generation of quantum annealing architecture, the connectivity of the shift register has been modified to reduce the total length by a factor of nearly three. The reduction in shift register length significantly reduces the time required to read out a single qubit. The three-phase operation of the shift register is described in detail here~\cite{Whittaker2016}. The speed at which data moves is set by the bandwidth of the QFP bias lines. For the current architecture we continue to use 30 MHz lines which sets the shift register data rate for each track to $\sim$ 10 Mbits/s.

\subsection{Frequency Multiplexed Resonator Readout}

At each end of every linear horizontal and vertical track (32 in total) we place a flux sensitive superconducting microresonator~\cite{Whittaker2016} (FASTR). The resonant frequency of each microresonator is set by an LC tank circuit. Part of the resonator inductance is provided by a direct current SQUID (DC-SQUID) loop that is coupled to the end stage of the QFP shift register track. Data in the last stage QFP body (circulating or countercirculating persistent current) modulates the inductance of the DC-SQUID loop which modulates the microresonator resonance frequency.

Each of the microresonators is connected to one of two microwave transmission lines that follow the perimeter of the processor.  We play a frequency tone that addresses a particular microresonator and monitor the transmission of this tone. The data state of the shift register modulates the microresonator frequency and thus the transmission of this tone. We can thus quickly read out the state of the shift register via a transmission measurement. The microresonators are separated in frequency and this frequency multiplexing allows us to read out all the microresonators in parallel.

\section{Floor Planning}
\label{sec:floorplan}

To simplify design and maintenance of our circuits we paid particular attention to designing our circuits in a modular fashion. Moreover, this enables experimentation with different device designs, degrees of connectivity, topologies, or even variations of qubit types in future processor generations.

We kept the notion of a plaquette from our earlier generation {\em (Picture of a square plaquette, cf. Fig. 7 in \cite{Bunyk2014})}, consisting of a square cluster of \fdac\ sources, surrounded by qubit and coupler analog circuitry. The latter consists of various SQUID loops, magnetically biased, from one side, by \fdac\ output transformers, and, from the other side, global room-temperature control lines. Each of these portions is well-isolated from the others and the environment by placing them into individual superconducting shielding boxes. All such devices have been designed to fit together tightly, facilitated by relative malleability of \fdac\ output transformers, which are necessary and substantial in area.  

The modularity of this approach results in streamlined design process. It achieves some degree of circuit optimality by mostly eliminating unused areas, allows for relatively easy design iterations, and fewer opportunities for crosstalks between nearby devices by re-using proven designs.  Over the course of this project, we made full designs of several processors with different connectivities and topologies, and we anticipate reusing very similar building blocks for our next-generation processors.

\section{Conclusion}
\label{sec:conclusion}

Performance studies on previous generations of quantum annealing technology have provided clear design guidelines for subsequent processor architectures. Design and fabrication innovations have allowed our current architecture to greatly expand the per-qubit connectivity without a substantial sacrifice in Hamiltonian energy scale. These innovations have kept processor programming and readout latencies fixed while reducing Hamiltonian specification errors. Future large-scale quantum annealing architectures will similarly focus on expanding connectivity, energy scale, and Hamiltonian fidelity. 

\bibliography{architecture_gen3}
\bibliographystyle{ieeetr}

\end{document}